\documentclass{ws-procs9x6}  
\usepackage{graphicx}

\newcommand{\ysur}{Y_{\rm S}}

\newcommand{\rcz}{R_{\rm CZ}}

\newcommand{\phib}{\Phi({\rm ^8B})}
\newcommand{\phibe}{\Phi({\rm ^7Be})}

\begin{document}
\title{An updated discussion of the solar abundance problem}

\author{F.L. Villante$^{1,2}$
}

\address{
$^{1}$Dipartimento di Scienze Fisiche e Chimiche,\\ Universit\`a
degli studi dell'Aquila,\\
Via Vetoio, L'Aquila, I-67100, Italy\\}
\address{
$^{2}$Laboratori Nazionali del Gran Sasso (LNGS),\\
Istituto Nazionale di Fisica Nucleare (INFN),\\
Via G. Acitelli 22,  Assergi  (AQ), I-67100, Italy
}

\author{A. Serenelli$^{3,4}$
}
\address{
$^{3}$Institute of Space Sciences (ICE, CSIC),\\
Carrer de Can Magrans S/N, Barcelona, E-08193, Spain
}
\address{
$^{4}$Institut d'Estudis Espacials de Catalunya (IEEC), \\
C/Gran Capita, 2-4, Barcelona, E-08034, Spain
}

%

\begin{abstract}
We discuss the level of agreement of  a new generation of standard
solar models (SSMs), Barcelona 2016 or B16 for short, with helioseismic
and solar neutrino data, confirming that models implementing the 
AGSS09met surface abundances, based on refined three-dimensional
hydrodynamical simulations of the solar atmosphere, 
do not not reproduce helioseismic constraints. 
We clarify that this solar abundance problem can be equally solved by a
change of the composition and/or of the opacity of the solar plasma,
since effects produced by variations of metal abundances are
equivalent to those produced by suitable modifications of the solar
opacity profile.
We discuss the importance of neutrinos produced in the CNO cycle  for removing the
composition-opacity degeneracy and the perspectives for their future detection.
\end{abstract}

\section{Introduction}

In the last three decades, there was an enormous progress in our
understanding of the Sun. The predictions of the Standard Solar Model
(SSM), which is the fundamental theoretical tool to investigate the
solar interior, have been tested by solar neutrino experiments and by
helioseismology. The deficit of the observed solar neutrino fluxes,
reported initially by Homestake\cite{Homestake1,Homestake2} and then confirmed by
GALLEX\cite{Gallex}, SAGE\cite{SAGE}, GNO\cite{GNO}, Kamiokande\cite{Kamiokande} and Super-Kamiokande\cite{SK}, generated the
so-called “solar neutrino problem”  which stimulated a deep investigation of the solar structure. 
The problem was solved in 2002 when the SNO experiment\cite{SNO}
obtained a direct evidence for flavour oscillations of solar neutrinos
and, moreover, confirmed the SSM prediction of the $^{8}{\rm B}$ neutrino flux.

Nowadays, we have a good knowledge of the solar neutrino oscillation
probability and a direct experimental determination of most of the
solar neutrino components. 
%
Super-Kamiokande\cite{SKLatest} and SNO\cite{SNOLatest} have provided a high accuracy determination of $^8{\rm B}$ neutrinos.
Borexino\cite{BorexPP,BorexLatest,Borex8BLatest} has recently obtained a direct measure of the {\em pp},
{\em pep}, $^{7}{\rm Be}$ and $^{8}{\rm B}$ solar neutrino fluxes
and it also has the potential to provide the first direct
measurements of the CNO neutrinos\cite{noiCNO} in the next future.
%
In addition, helioseismic observations have allowed to determine
precisely  several important properties of the Sun, such as the depth
of the  convective envelope which is known at the $\sim 0.2\%$ level,
the surface  helium abundance which is obtained at the $\sim 1.5\%$
level  and the sound speed profile which is determined with an
accuracy  equal to $\sim 0.1\%$ in a large part of the Sun (see
e.g. Refs.\citenum{HelioReview1} and \citenum{HelioReview2} and
J. Christensen-Dalsgaard contribution to these proceedings.  As a results of these
observations,  the solar structure is now very well constrained, so
that the Sun  can be used as a solid benchmark for stellar evolution
and  as a “laboratory” for fundamental physics.

A new solar problem has, however, emerged during the last
years. Recent determinations of the  photospheric heavy element
abundances\cite{agss09,caffau,ags05} indicate  that the Sun metallicity is lower than
previously assumed\cite{gs98,gn93}.  Solar models that incorporate these lower
abundances  are no more able to reproduce the helioseismic results. As
an example,  the sound speed predicted by SSMs at the bottom of the
convective envelope  disagrees at the $\sim 1\%$ level with the value
inferred by  helioseismic data (see e.g. Ref.\citenum{serenelli14}). Detailed studies have
been done to resolve  this controversy (see e.g. Ref.\citenum{Basu:2007fp}), but a
definitive  solution of this solar abundance problem still has to
be obtained.

 In this review, we provide a quantitative discussion of the solar
 abundance problem. The plan of the paper is as follows. 
In Secs.\ref{sec:SSM} and \ref{sec:B16}, we present a a new generation of standard
solar models (SSMs), Barcelona 2016 or B16 for short, and we quantify
the level of agreement of models implementing different surface
abundances with helioseismic and solar neutrino data.
In Sec.\ref{sec:composition}, we discuss the degeneracy between the
effects produced by a modification of the radiative opacity and of
those produced by a modification of the
surface composition.
In Sec.\ref{sec:CNO}, we dicsuss the importance of neutrinos
produced in the CNO cycle for removing the composition-opacity
degeneracy and for solving the solar abundance problem.
Finally, we summarise our results in Sec.\ref{sec:summary}.


\section{The B16 standard solar models}
\label{sec:SSM}

SSMs are a snapshot in the evolution of a 1$M_\odot$ star, calibrated to match present-day surface properties of the Sun.  The calibration is done by adjusting the mixing length parameter ($\alpha_{\rm MLT}$) and the initial helium and metal mass fractions ($Y_{\rm ini}$ and $Z_{\rm  ini}$ respectively) in order to satisfy the constraints imposed by
the present-day solar luminosity $L_{\odot}$, radius $R_\odot$, and surface metal to hydrogen abundance ratio $(Z/X)_\odot$.
The new B16 models share with previous  calculations\cite{serenelli11} much of the input physics, but include important updates. A brief account of few relevant ingredients is given in
the following.\\

{\bf Equation of State:}  B16 SSMs employ, for the first time, EoS tables calculated consistently for each of the compositions used in the solar
calibrations by using FreeEOS\cite{freeEOS}. \\

{\bf Nuclear rates:} 
The rates of ${\rm p(p,e^{+}\nu_e)d}$,  ${\rm ^{\rm 7} Be(p,\gamma)^{\rm 8}B}$ and 
${\rm ^{\rm 14}N(p,\gamma)^{\rm 15}O}$ reactions have been updated, see Tab.\ref{tab:supdated}\footnote{For the ${\rm p(p,e^{+}\nu_e)d}$ reaction, the quoted value for
$S_{11}(0)$ underestimates the actual increase of the rate 
because the variation of $S_{11}(E)$ at solar energies
is dominated by changes in the first and higher order
derivatives of the Taylor expansion of the astrophysical factor 
around $E=0$ (see Ref. \citenum{noi} for details).}.
For the important reaction $^{3}{\rm He}(^{4}{\rm  He},\gamma)^{7}{\rm
  Be}$ (not included in Tab.\ref{tab:supdated}), two recent  analyses\cite{deboer,iliadis} 
have provided determinations of the astrophysical factor 
that differs by about $6\%$ (to be compared with a claimed
accuracy equal to $4\%$ and $2\%$ for Refs. \citenum{deboer} and \citenum{iliadis}, respectively).
Considering that the results from Refs.\citenum{deboer} and \citenum{iliadis}
bracket the previously adopted value from Ref.\citenum{adelberger11},  
the latter was considered as preferred choice in B16 SSMs.\\

\begin{table}[t]
\tbl{Astrophysical S-factors (in units of MeV b) and uncertainties updated in this work. Fractional changes with respect to Ref.\citenum{adelberger11} are also included.}
{\begin{tabular}{c | c c c c}
\toprule
         & S(0)  & Uncert.(\%) & $\Delta S(0)/S(0)$ & Ref. \\
\hline
$\rm{S_{11}}$ & $4.03 \cdot 10^{-25}$& 1 & 0.5\% & \citenum{s11a,s11b,s11c}\\
$\rm{S_{17}}$ & $2.13 \cdot 10^{-5}$ & 4.7  & +2.4\% & \citenum{s17}\\
$\rm{S_{114}} $ & $1.59 \cdot 10^{-3}$& 7.5 & -4.2\% & \citenum{s114}\\
\botrule
\end{tabular}}
\label{tab:supdated}
\end{table}

{\bf Radiative opacities:} In Ref.\citenum{serenelli11} the opacity error was
modelled as a 2.5\% constant factor at 1$\sigma$ level, comparable to
the maximum difference between OP \cite{OP} and OPAL \cite{OPAL} 
opacities in the solar radiative region.  It was shown, however, in Ref.\citenum{villanteOPA}
that this prescription underestimates the contribution of opacity
uncertainty to the sound speed and convective radius error budgets 
because the effects produced by opacity variations in different zones of
the Sun compensate among each other and integrate to zero for a
global rescaling of the opacity. 
Moreover this is not realistic because the accuracy of opacity
calculations is expected to be better at the solar core 
than in the region around the 
base of the convective envelope.
Taking this into account, the following
parameterization for the opacity change $\delta \kappa (T)$ was considered:
\begin{equation}
\delta \kappa(T) = a + b \frac{\log\left(T_{\rm C}/T\right)}{\Delta}
\end{equation}
where $T$ is the temperature of the solar plasma, $\Delta =
\log\left(T_{\rm C}/T_{\rm CZ}\right) = 0.9$, $T_{\rm C} = 15.6 \times
10^6\, {\rm K}$ and $T_{\rm CZ} = 2.3 \times 10^6\, {\rm K}$ 
are the temperatures at the solar center and at the bottom of the
convective zone respectively. The parameters $a$ and $b$
are treated as independent random variables with mean
equal to zero and dispersions $\sigma_a = 2\%$ and $\sigma_b = 6.7\%$,
respectively.  This corresponds to assuming that the opacity error at the solar
center is $\sigma_{\rm in} =\sigma_a = 2\% $ , while it is given by 
$\sigma_{\rm out} =(\sigma_a^2 + \sigma_b^2)^{1/2} = 7\%$ at the base of the
convective zone, as can be motivated by the recent experimental results of
Ref.\citenum{OPAexp} and the theoretical work by
Ref.\citenum{OPAtheo}.\\

{\bf Surface composition:} 
The solar surface composition is a fundamental constraint in the construction of SSMs.
In this paper, we consider two different canonical sets of
solar abundances which are the same employed in Ref.\citenum{serenelli11}: 
\begin{itemize}
\item{GS98 - }
Photospheric (volatiles) + meteoritic (refractories)
abundances from Ref.\citenum{gs98} 
that correspond to metal-to-hydrogen ratio used for the calibration
$(Z/X)_\odot = 0.0229$; 
\item{AGSS09met - }
Photospheric (volatiles) + meteoritic
(refractories) abundances from Ref.\citenum{agss09} that give  
$(Z/X)_\odot = 0.0178$.
\end{itemize}
Note that the recent results from Refs.\citenum{agss15a,agss15b,agss15c} that have updated the
abundances of Ref.\citenum{agss09} for all but CNO elements (which are the
most abundant among the volatiles elements) do not lead to a
revision of the AGSS09met composition.

\begin{table}[t]
\tbl{Neutrino fluxes for the two B16 SSMs and 
 as determined by Ref.\citenum{bergstrom}. The fluxes are given in units of 
$10^{10}\,(\rm{pp})$, $10^{9}\,(\rm{^7 Be})$, $10^8\,(\rm{pep,^{13}N,^{15}O})$,
 $10^6\,(\rm{^8B,^{17}F})$ and $10^3\,(\rm{hep})$ $\rm{cm^{-2} s^{-1}}$. 
The last two lines give the surface helium $Y_{\rm S}$ and the
convective radius $R_{\rm CZ}$. The observational values are given by
Refs. \citenum{YsErr} and \citenum{RczErr}, respectively.}
{\begin{tabular}{c c c c}
\toprule
 & GS98 &AGSS09met & Obs\\
\hline
$\Phi ({\rm pp})$ & $5.98(1 \pm 0.006)$ & $6.03(1 \pm 0.005) $&$5.971^{+0.037}_{-0.033}$\\
$\Phi ({\rm pep})$  &$ 1.44(1 \pm 0.01) $&$1.46(1 \pm 0.009) $&$1.448 \pm 0.013$\\
$\Phi ({\rm hep})$ & $7.98(1 \pm 0.30) $&$8.25(1 \pm 0.30) $&$ 19^{+12}_{-9}$\\
$\Phi (^7{\rm Be})$ &$ 4.93(1 \pm 0.06)$ &$4.50(1 \pm 0.06) $&$4.80^{+0.24}_{-0.22}$\\
$\Phi (^8{\rm B})$ & $5.46(1 \pm 0.12)$ &$4.50(1  \pm 0.12) $&$5.16^{+0.13}_{-0.09}$\\
$\Phi (^{13}{\rm N})$ & $2.78(1 \pm 0.15)$ &$2.04(1  \pm  0.14) $&$\le 13.7$\\
$\Phi (^{15}{\rm O})$ & $2.05(1 \pm 0.17)$ &$1.44(1 \pm 0.16) $&$\le 2.8$\\
$\Phi (^{17}{\rm F})$ & $5.29(1 \pm 0.20)$ &$3.26(1 \pm 0.18) $&$\le 85$\\
\hline
$Y_{\rm S}$ & $0.2426\pm0.0059$ &$0.2317\pm0.0059$&$0.2485\pm0.0035$\\
$R_{\rm CZ}$ & $0.7116\pm0.0048$ &$0.7223\pm0.0053$&$0.713\pm0.001$\\
\botrule
\end{tabular}}
\label{tab2}
\end{table}

\section{B16-SSMs results}
\label{sec:B16}

\begin{figure}[t]
	\centering
	\includegraphics[width=0.7\columnwidth]{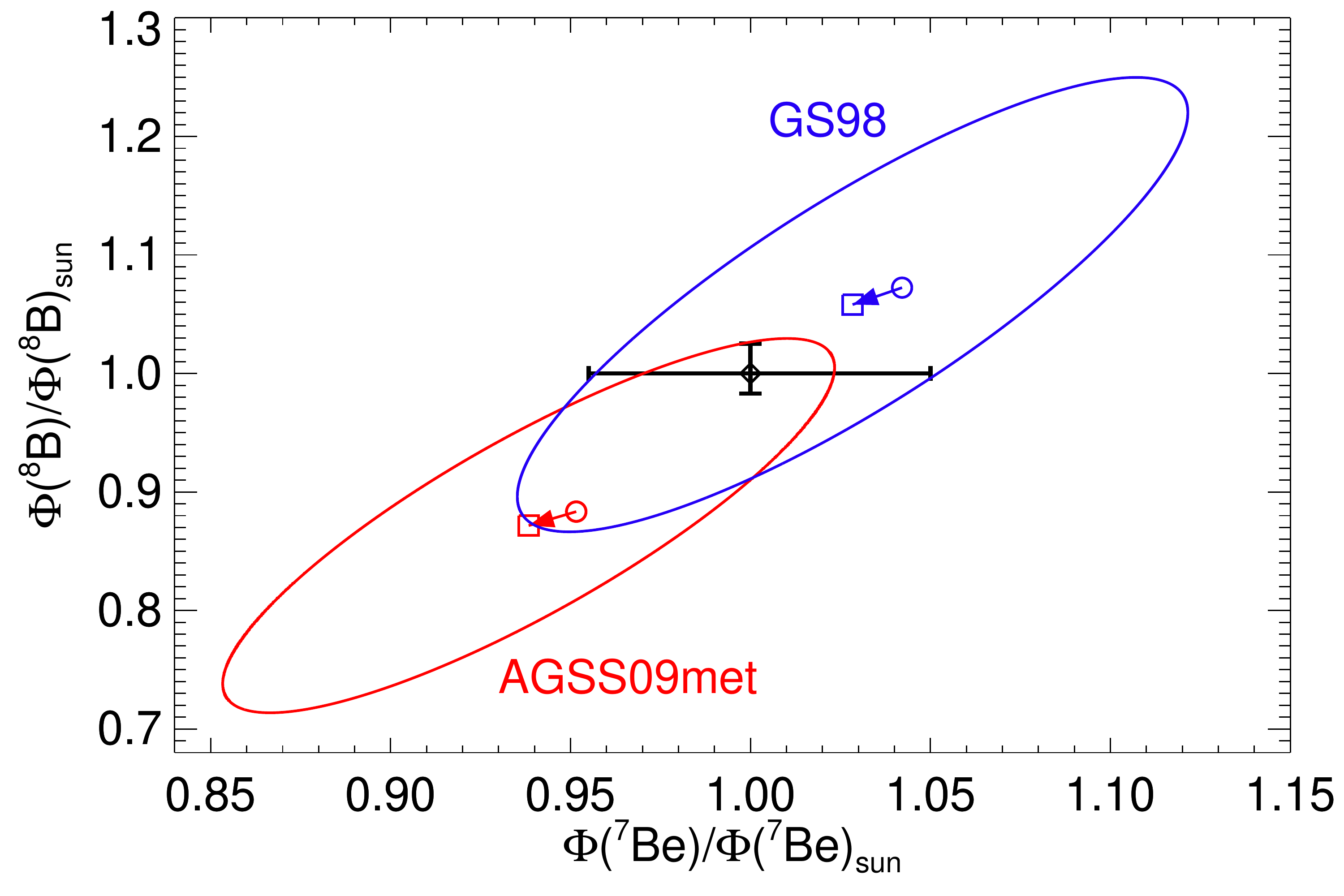}
        \caption{\small\em $\phib$ and $\phibe$ fluxes normalized to
          the solar values obtained in Ref.\citenum{bergstrom}. Black circle and error bars:
          solar values. Squares and circles: results for B16 (current)
          and (older) generation of  SSMs respectively. Ellipses denote theoretical 1$\sigma$ C.L. for 2 dof. \label{fig1}}
\end{figure}

\begin{figure}[t]
	\centering 
	\includegraphics[width=0.7\columnwidth]{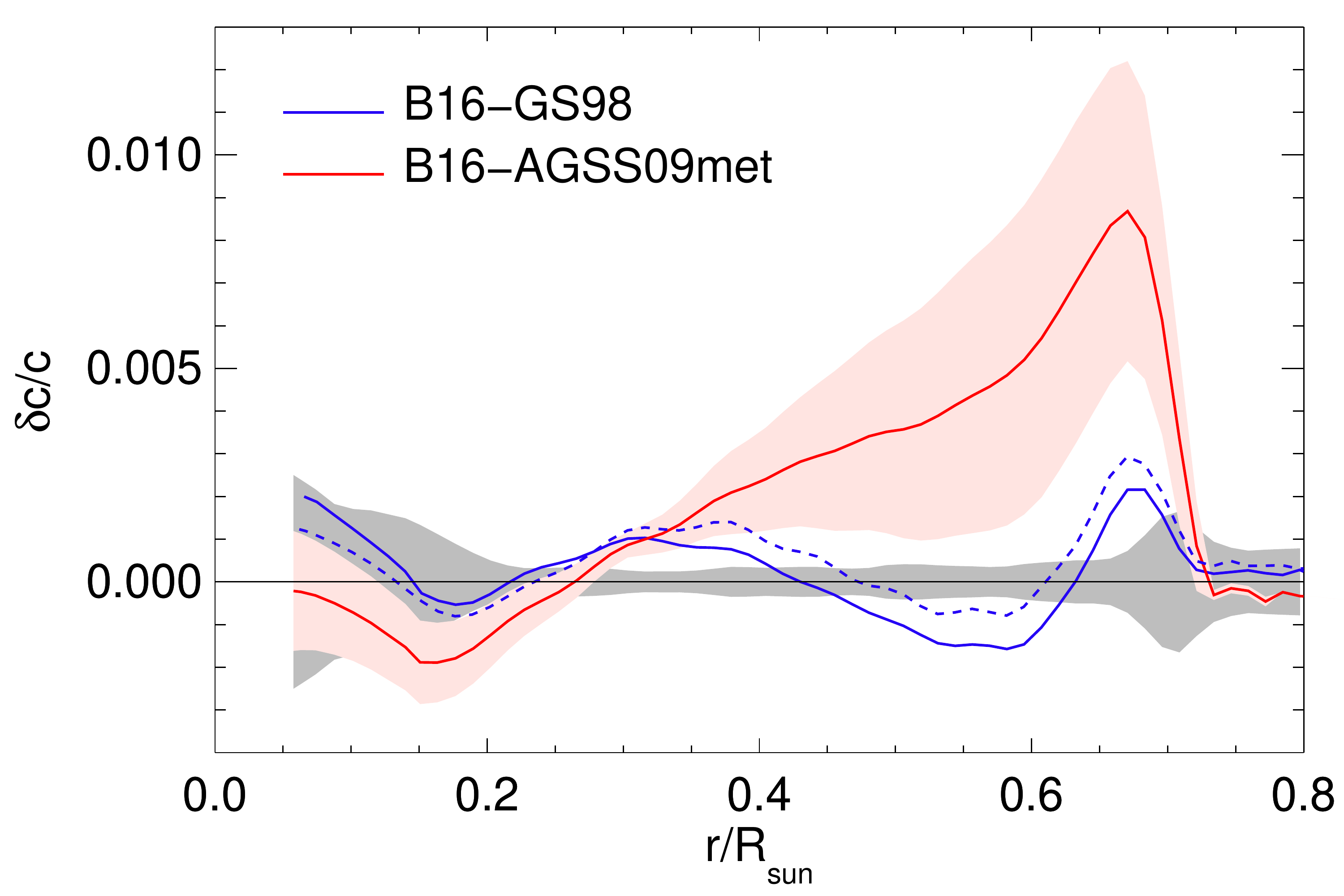}
        \caption{\small\em Fractional sound speed difference $\delta
          {\rm c /c = (c_\odot - c_{\rm SSM})/c_{\rm SSM}}$. The grey
          shaded region corresponds to errors in the helioseismic
          inversion procedure. The red shaded region around the
          AGSS09met central value (solid red line) describes uncertainties
          in SSM calculations. An equivalent
          relative error band holds around the central value of the
          GS98 central value (solid blue line) which we do not plot
          for the sake of clarity. Dashed line shows, for comparison,
          results for old SSM calculations \cite{serenelli11}. \label{fig2}}
\end{figure}

The main results obtained with the new generation of B16 SSMs for the
two choices of solar composition, GS98 and AGSS09met, are shown in 
Tab.\ref{tab2}, Fig.\ref{fig1} and \ref{fig2} and are discussed below.
In Tab. \ref{tab3} we quantify the level of agreement between the B16-SSMs 
predictions and different ensembles of solar observables. The $\chi^2$
values reported in Tab.\ref{tab3} are calculated 
by including experimental and theoretical (correlated) uncertainties 
as described in Refs.\citenum{villanteCC} and \citenum{noi}.
\\

{\bf Neutrino fluxes:} The updates of nuclear reaction rates have a direct effect
on neutrino production. 
In particular, the boron and beryllium neutrino fluxes are reduced for both GS98 and AGSS09met compositions
by about 2\% with respect to previous SSM calculations \cite{serenelli11}. 
The overall reduction in the $\Phi(^{8} {\rm B})$ and $\Phi(^{8} {\rm  Be})$ fluxes 
comes from the increase in $S_{11}$. In the case of $\Phi(^{8} {\rm  B})$, 
this is partially compensated by the 2.4\%  increase in $S_{17}$.
The most important changes in the neutrino fluxes occur for
$\Phi(^{13}{\rm N})$ and $\Phi(^{15}{\rm O})$, in the CN-cycle. 
The expectation values in the B16 SSMs are about 6\% and 8\%  lower
than for the previous SSMs \cite{serenelli11}. This results from the combined
changes in the p+p and $^{14}$N+p reaction rates.

The predicted fluxes should be compared with the observational values in the
last column of Tab.\ref{tab2} which have been obtained in
Ref.\citenum{bergstrom} from a fit to the results of solar
neutrino experiments by allowing for three-flavour neutrino oscillations.
%
Note that observational errors for $\Phi(^{8} {\rm B})$
and $\Phi(^{8} {\rm  Be})$ fluxes are smaller than uncertainties in
theoretical predictions, as can be also appreciated in Fig.\ref{fig1} 
where we summarize the present situation for these
two components of the solar neutrino spectrum.
On the contrary, CN fluxes have not yet been determined experimentally 
and the global analysis of solar neutrino data  provides only the upper
limits included in Tab.\ref{tab2}.
%

From the quantitative comparison of B16-GS98 and B16-AGSS09met predictions 
with the experimentally inferred neutrino fluxes, we conclude that both
calculations are consistent with neutrino data within $1\sigma$.
We obtain indeed $\chi^2/{\rm dof}\lesssim 1$  for both assumed compositions
when considering $\Phi(^{7} {\rm Be}) + \Phi(^{8} {\rm  B})$ and/or ``all $\nu$-fluxes'' experimental
deteminations, as it is reported in Tab.\ref{tab3}.\\

{\bf Helioseismology:} 
In the last two lines of Tab.\ref{tab2}, we report two helioseismic quantities widely used in assessing the
quality of SSMs, i.e. the surface helium abundance $Y_{\rm S}$ and the depth of the convective
envelope $R_{\rm CZ}$, together with the corresponding seismically determined
values. The model errors associated to these quantities are larger
than previously computed because of the different treatment 
of uncertainties in radiative opacities. Compared to previous
SSMs\cite{serenelli11},  we find a small decrease in the predicted $Y_{\rm S}$ by 0.0003
and in the predicted $R_{\rm CZ}$ by $0.0007 R_\odot$ for both
compositions.  These small changes together with the
larger theoretical uncertainties lead B16-GS98 to a 0.9$\sigma$ ($Y_{\rm S}$) and
0.3$\sigma$ ($R_{\rm CZ}$) difference with respect to data while for B16-AGSS09met
differences are at the 2.5$\sigma$ ($Y_{\rm S}$) and 1.7$\sigma$
($R_{\rm CZ}$) level. When combined together, see Tab.\ref{tab3}, these helioseismic
observables give a $\chi^2$ (for 2 dof) equal to 0.9 (6.5) for B16-GS98 (B16-AGSS09met) predictions,
showing a preference for the high metallicity assumption.

Finally, Fig.\ref{fig2} shows the fractional difference between the sound
speed inferred from helioseismic frequencies and that predicted by
B16 SSMs as a function of solar radius for the two choices of solar composition. 
The solar sound speed has been obtained by new inversions based on the
so-called BiSON-13 dataset \cite{BISON} and using consistently both B16
SSMs as reference models. Results are only slightly different with
respect to previous calculations, mainly as a result of the updated
$S_{11}(0)$ value. The red shaded region around the AGSS09met central
value (solid red line) describes theoretical uncertainties in SSM calculations. An
equivalent relative error band holds around the central value of the
GS98 central value (solid blue line) which we do not plot for the sake of clarity.
The sound speed profile is also affected by observational errors which
are due to uncertainties in the measured helioseismic frequencies, numerical parameters inherent to the inversion
procedure and the solar model used as a reference model for performing
the inversion. These errors have been evaluated as described in Refs.\citenum{HelioReview1} and \citenum{noi}
and correspond to the grey shaded region in Fig.\ref{fig2}

We see that B16-GS98 model yields a much better agreement, everywhere in the solar structure, 
with the helioseismically derived sound speed profile than
B16-AGSS09met. In particular,  the B16-AGSS09 model disagrees by $\sim  1\%$
with sound speed inferred from helioseismology at the bottom of the convective 
envelope. This has to be compared with a theoretical uncertainty of $\sim
0.3\%$ and an error in the inversion procedure smaller than $0.1\%$. 
Using theoretical and experimental uncertainties as described in Ref.\citenum{villanteCC},
we can compare how well the predicted sound speed profiles of B16-GS98
and B16-AGSS09met agree with helioseismic inferences. 
For this, we use the same 30 radial points employed in
Refs.\citenum{villanteCC}. 
Results are shown in the second row of Tab.\ref{tab3}. For 30
degrees-of-freedom (dof),  B16-GS98 gives $\chi^2 = 58$, or a $3.2\sigma$ agreement with data.
For B16-AGSS09met results are $\chi^2 = 76.1$, or $4.5\sigma$.
It is important to notice that in the case of B16-GS98, the largest
contribution to the sound speed $\chi^2$ comes from the narrow region 
$0.65 < r/R_\odot < 0.70$ that comprises 2 out of all the 30 points. 
If these two points are removed from the analysis $\chi^2$ is reduced
from 58 to 34.7, equivalent to a $1.4\sigma$ 
agreement with the solar sound speed (entry identified as $\delta c/c$
no-peak in Tab.\ref{tab3}). 
For B16-AGSS09met this test leads to a $2.7\sigma$ result. 
This exercise highlights the qualitative difference between SSMs with
different compositions; it shows that for GS98 the problem is highly
localized whereas for AGSS09met the disagreement between SSMs and
solar data occurs at a global scale, i.e. the solar abundance problem.

\begin{table}[t]
\tbl{Comparison of B16 SSMs against different ensembles of solar observables.}
{\begin{tabular}{cccccc}
\toprule
\multicolumn{2}{c}{} &   \multicolumn{2}{c}{GS98} & \multicolumn{2}{c}{AGSS09met}\\ \hline
Case & \multicolumn{1}{|c|}{dof}  &  $\chi^2$ & p-value\,$(\sigma)$ & $\chi^2$ & p-value\,$(\sigma)$ \\ \hline
$\ysur + \rcz$ only & \multicolumn{1}{|c|}{2} & 0.9 & 0.5 & 6.5 & 2.1 \\
$\delta c/c$ only  & \multicolumn{1}{|c|}{30} & 58.0 & 3.2 & 76.1 & 4.5 \\ 
$\delta c/c$ no-peak  & \multicolumn{1}{|c|}{28} & 34.7 & 1.4 & 50.0 & 2.7 \\ 
$\phibe+\phib$  & \multicolumn{1}{|c|}{2} & 0.2 & 0.3 & 1.5 & 0.6 \\
all $\nu$-fluxes  & \multicolumn{1}{|c|}{8} & 6.0 & 0.5 & 7.0 & 0.6 \\
\hline
global & \multicolumn{1}{|c|}{40} & 65.0 & 2.7 & 94.2 & 4.7 \\
global no-peak & \multicolumn{1}{|c|}{38} & 40.5 & 0.9 & 67.2 & 3.0 \\
\botrule
\end{tabular}}
\label{tab3}
\end{table}

\section{The opacity-composition degeneracy}
\label{sec:composition}

The interpretation of the solar abundance problem is complicated by
the degeneracy between effects produced by a modification of the radiative
opacity $\kappa(\rho,T,Y,Z_{\rm i})$ and effects induced by a change of the heavy
element admixture $\{z_{\rm i}\}$, expressed here in terms of the
quantities  $z_{\rm i} \equiv Z_{\rm i,b}/X_{\rm b}$ where $Z_{\rm i,b}$ is
the surface abundance of the $i$-element and $X_{\rm b}$ is that of hydrogen.

This degeneracy was discussed in quantitative terms in Ref.\citenum{villanteOPA}
by using the linear solar model (LSM) approach introduced in
Ref.\citenum{villanteLSM}. 
By neglecting the role of metals in the equation of state
and in the energy generation coefficient, it was shown that the source term  $\delta \kappa(r)$ that 
drives the modification of the solar properties
and that can be constrained by observational data 
can be written as the sum of two contributions:
\begin{equation}
\delta \kappa(r) = \delta \kappa_{\rm I}(r) + \delta \kappa_{\rm Z}(r)
\label{kappasource}
\end{equation}
The first term $\delta \kappa_{\rm I}(r)$, which we refer to as {\em intrinsic} opacity change, 
represents the fractional variation of the opacity  along the solar profile and it is given by:
\begin{equation}
\label{kappaintrinsic}
\delta \kappa_{\rm I}(r) = \frac{\kappa(\overline{\rho}(r),\overline{T}(r),\overline{Y}(r),\overline{Z}_{i}(r))}
{\overline{\kappa}(\overline{\rho}(r),\overline{T}(r),\overline{Y}(r),\overline{Z}_{i}(r))} -1
\end{equation}
where the notation $\overline {Q}$ indicates, here and in the
following, the value for the generic quantity $Q$ that is obtained in a
reference SSM calculation. 
This contribution is obtained when we revise the opacity function $\kappa(\rho,T,Y,Z_{\rm i})$
and/or we introduce new effects, like e.g. the accumulation of few GeVs WIMPs in the solar 
core that mimics a decrease of the opacity at the solar center,  see
e.g. Ref.\citenum{villanteWIMPs} and references therein.
The second term $\delta \kappa_{\rm Z}(r)$, 
which we refer to as {\em composition} opacity change, 
describes the effects of a variation of $\left\{ z_{\rm i} \right\}$.
It takes into account that a modification of the photospheric admixture implies a different 
distribution of metals inside the Sun and, thus, a different opacity profile, even if the function
$\kappa(\rho,T,Y,Z_{\rm i})$ is unchanged. The contribution $\delta \kappa_{\rm Z}(r)$ is given by:
\begin{equation}
\delta \kappa_{\rm Z}(r) = \frac{
\overline{\kappa}(\overline{\rho}(r),\overline{T}(r),\overline{Y}(r),Z_{i}(r))}
{\overline{\kappa}(\overline{\rho}(r),\overline{T}(r),\overline{Y}(r),\overline{Z}_{i}(r))} - 1 
\end{equation}
where $Z_{\rm i}(r) \simeq \overline{Z}_{\rm i}(r) \, (z_{\rm i} / \overline{z}_{\rm i})$ and can be calculated as:
\begin{equation}
\label{kappacomposition}
\delta \kappa_{\rm Z}(r)  \simeq \sum_{i} \left.\frac{\partial \ln \overline {\kappa}}{\partial \ln Z_{i}} \right|_{\rm SSM} \; \delta z_{\rm i,b}
\end{equation} 
where $\delta z_{\rm i}$ represents the fractional variation of $z_{\rm i}$  and 
the symbol $|_{\rm SSM}$ indicates that we calculate the derivatives along the density, temperature and 
chemical composition profiles predicted by the reference SSM.
Equation (\ref{kappasource}), although being approximate, is quite
useful because it  makes explicit the connection (and the degeneracy)
between the effects produced by a modification of the radiative opacity and of those produced by a modification of the heavy element admixture.

The response of the Sun to an arbitrary modification of the opacity
$\delta  \kappa(r)$ was studied in Ref.\citenum{villanteOPA} (see also
Refs.\citenum{JCD} and \citenum{noi}) by computing numerically the kernels that, in a linear approximation, relate the opacity change 
$\delta \kappa(r)$ to the corresponding modifications of the solar observable properties. 
The following quantities where considered: the sound speed profile,
the surface helium abundance, the inner boundary of the convective envelope, 
the solar neutrino fluxes, the central solar temperature.
It was shown that different observable quantities probe different regions of the Sun. 
Moreover, effects produced by variations of opacity in distinct zones of the Sun may compensate among 
each other. In this respect, it was noted that {\em the sound speed profile and the depth of the convective 
envelope are practically insensitive to a global rescaling
of the opacity.}
As a consequence, the discrepancy between the helioseismic determinations
of these quantities and the predictions of SSMs implementing AGSS09met
composition
can be equally solved by a $\sim 15\%$ decrease of the opacity a the center of the Sun 
or by a $\sim 15 \%$ increase of the opacity in the external radiative region.
The degeneracy between these two possible solutions is, however, broken by the ``orthogonal''
information provided by the measurements of the surface helium abundance and of the
boron and beryllium neutrino fluxes, that fix the scale of opacity and indicate that
only the second possibility can effectively solve the solar
composition problem.

\begin{figure}[t]
	\centering 
	\includegraphics[width=0.7\columnwidth]{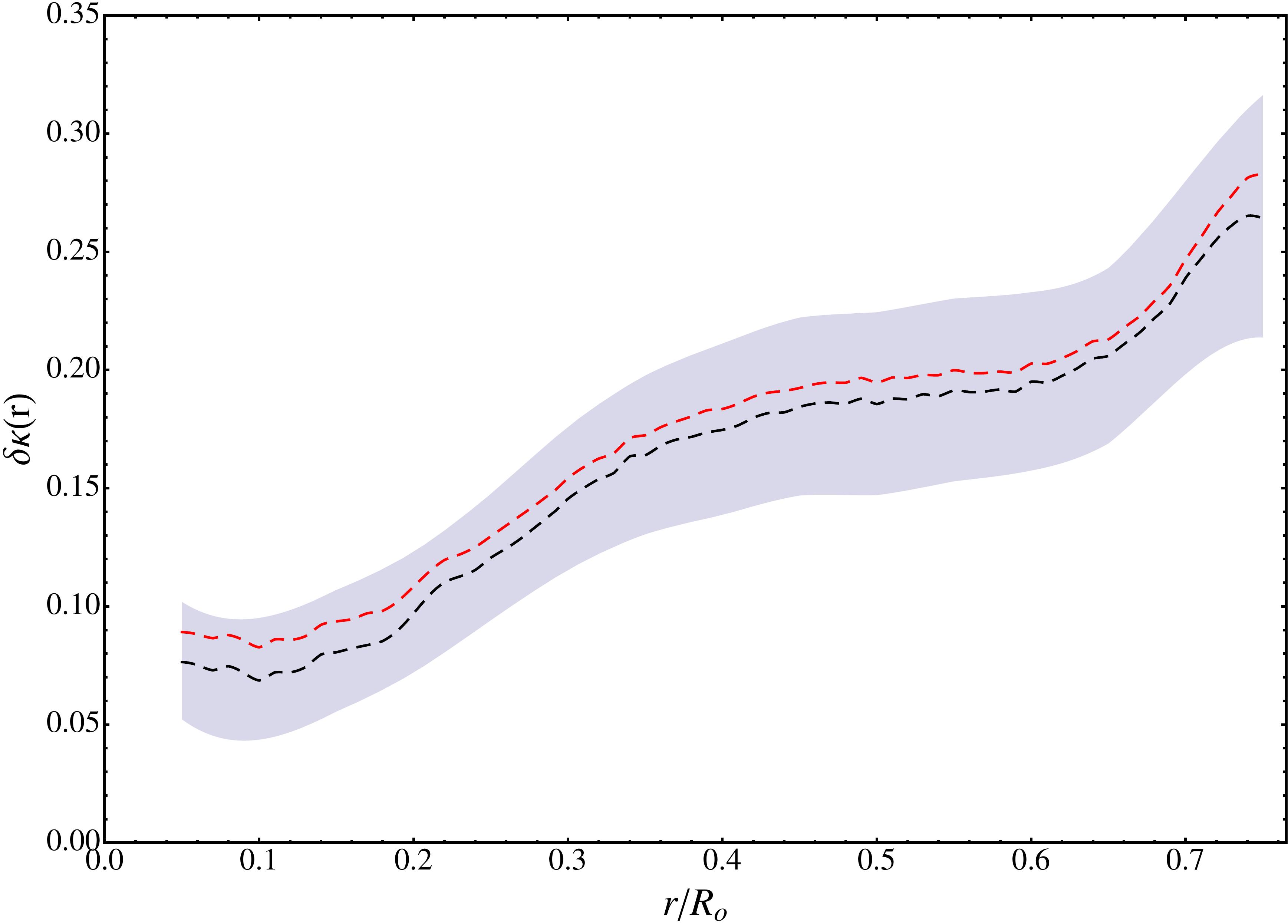}
        \caption{\small\em 
The  effective opacity change $\delta \kappa(r)$ of  solar  models
that  provide a  good  fit  to helioseismic and solar neutrino data 
when $\left(\delta z_{\rm  CNO},\delta z_{\rm Ne}, \delta z_{\rm met}\right)$ are  allowed to vary.  The black dashed line  correspond to
the best  fit model. The  red dashed line  correspond to the best  fit model  obtained  with  the additional  assumption  that  $\delta Z_{\rm  Ne}=\delta
  z_{\rm  CNO}$, i.e.  that the  neon-to-oxygen ratio  is equal  to
  the value  prescribed by AGSS09met compilation. See
  Ref.\citenum{villanteCC} for details.
\label{fig3}}
\end{figure} 

Incidentally, the above conclusion is confirmed by the analysis
presented in Ref.\citenum{villanteCC} where helioseismic and solar neutrino
data are used to infer the optimal composition of the Sun. 
The effective opacity change $\delta \kappa(r)$ (with respect to
reference SSM implementing AGSS09met surface composition and OP 
opacity\cite{OP}) 
that produces a good fit to observational data is shown in Fig.\ref{fig3}.
We see that $\delta \kappa(r)$ is well constrained by the available
observational information. 
Opacity should be increased by $\sim {\rm few}$ percent at the center of the Sun
and by $\sim 25\%$  at the bottom of the convective envelope, as it was
calculated by Ref\citenum{villanteOPA}. 
The red dashed line in Fig.\ref{fig3} is obtained in a two parameter
analysis in which elements are grouped as volatiles (i.e., C, N, O,
and Ne) and refractories (i.e., Mg, Si, S, and Fe).
The optimal surface composition is found by increasing the abundance
of volatiles by $(45 \pm 4)\%$ and that of refractories by $(19 \pm 3)\%$ 
with respect to the values provided by AGSS09met.
The black lines is obtained in a three parameter analysis in which the
neon-to-oxygen ratio is allowed to vary within the currently allowed
range (i.e., $\pm 30\%$ at $1\sigma$). The best-fit composition is obtained by
increasing by $(37 \pm 7)\%$ the CNO elements; by $(80 \pm 26)\%$ the neon;
and by $(13 \pm 5)\%$ the refractory elements.
We see that the two lines coincide at the 2\% level or better. From
this, we infer that the reconstructed opacity profile does not depend
on the assumed heavy element grouping. 
Moreover, we understand that the best-fit compositions obtained
in the two and three parameter analyses cannot be discriminated by the adopted observational constraints.

\section{CNO neutrinos}
\label{sec:CNO}

As it is well known, the degeneracy between radiative opacity and 
solar composition can be removed by measuring the neutrino fluxes produced in the CNO cycle. 
The peculiarity of the CNO cycle is that it uses
carbon, nitrogen and oxygen nuclei which are present in the core of the Sun as catalysts for H
burning. 
As a consequence, its contribution to neutrino and energy production, beside
depending on the solar temperature stratification (and thus on the
opacity profile of the Sun), is approximately proportional to the stellar-core 
number abundance of CNO elements.
This additional dependence can be used in combination with other
helioseismic and solar neutrino probes to obtain a direct
determination of the CNO core abundances.

This possibility was discussed on a quantitative
basis in Ref.\citenum{SerenelliHaxton}.  By taking advantage of the
fact that $\Phi(^{13}{\rm N})$ and $\Phi(^{15}{\rm O})$ fluxes,  which are produced by 
the dominant CN-branch of the CNO cycle, have a similar dependence 
on the core temperature of $^8$B neutrinos, it was shown that 
the measured $\Phi(^8{\rm B})$ flux can be used to largely eliminate
environmental uncertainties (solar age, opacity, luminosity,...)
affecting the CN fluxes predictions.
This permits to translate a future measurement of CN neutrinos into 
a determination of the carbon and nitrogen abundances in the
solar core with an accuracy that is sufficient to discriminate between
opacity and/or composition changes as the solution
of the solar abundance problem. 
Moreover, by directly comparing surface and core abundances
one could, in principle, test the standard chemical evolution paradigm, according
to which the Sun was born chemical homogenous and then it has evolved
its internal composition due to nuclear reactions and elemental
diffusion. 

\begin{figure}[t]
	\centering 
	\includegraphics[width=0.7\columnwidth]{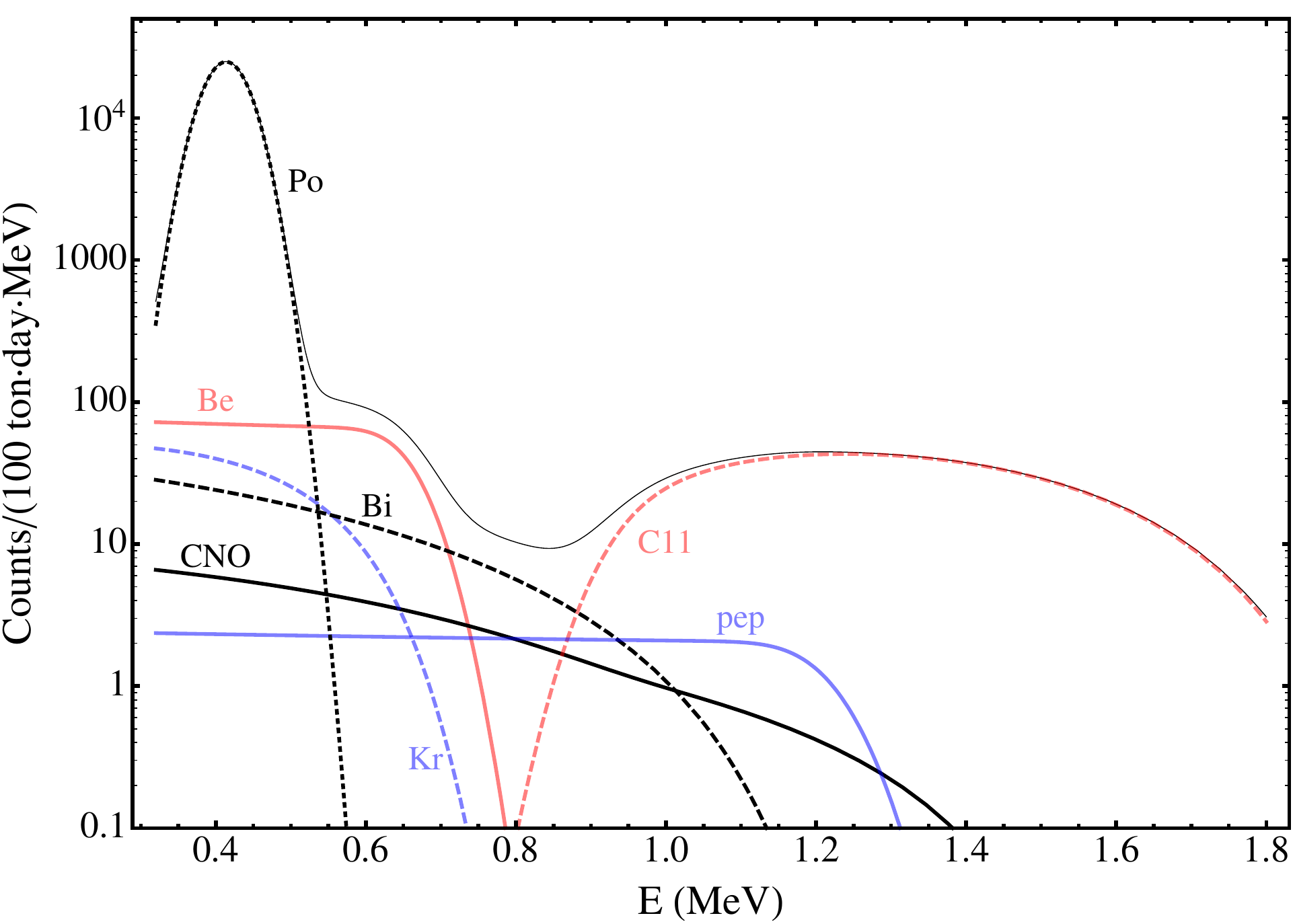}
        \caption{\small\em 
The expected event spectrum in solar neutrino liquid scintillator detectors
calculated by assuming background levels and detector energy resolution comparable to those obtained by 
Borexino detector \cite{PRLBorex}. See Ref.\citenum{noiCNO} for details.
\label{fig4}}
\end{figure}

At present, we still miss a direct observational evidence for CNO
energy generation in the Sun.  We only have a loose upper limit on CNO
neutrino fluxes  obtained by combining the results of the various
solar neutrino experiments. 
Direct detection of CNO neutrinos is a very difficult task. 
Not only the flux is relatively low, but also their energy is not large.  
The neutrinos produced by $\beta$-decay processes in the CNO cycle,
i.e.
\begin{eqnarray}
\nonumber
^{13}{\rm N}&\rightarrow&^{13}{\rm C}+e^++\nu_e\\
^{15}{\rm O}&\rightarrow&^{15}{\rm N}+e^++\nu_e\\
\nonumber
^{17}{\rm F}&\rightarrow&^{17}{\rm O}+e^++\nu_e
\end{eqnarray}
have continuous energy spectra with endpoints at about $\sim 1.5\,{
\rm MeV}$. 
Differently from the monochromatic $^{7}$Be and pep solar
neutrinos, they do not produce specific spectral features that permit
to extract them unambiguously from the background event spectrum in
high purity liquid scintillators. 
In particular, as it is shown in Fig.\ref{fig4}, the electrons produced by the
$\beta$-decay of Bismuth-210 to Polonium-210 have a spectrum that is
similar to that produced by CNO neutrinos. 
As a consequence, spectral fits are able to determine only combined “Bismuth+CNO” contribution, as it is done e.g. by Borexino\cite{BorexBiCNO}.

In order to remove this degeneracy, a method to determine the Bi-210 decay rate which is based on the
relationship between the Bi-210 and Po-210 abundances was proposed in Ref.\citenum{noiCNO}. 
Polonium-210, which is the Bismuth-210 daughter, is unstable and
decays with a lifetime  $\tau_{\rm Po}\sim 200$ days emitting a monochromatic $\alpha$
particle that can be easily detected. In the absence of Bismuth-210,
the $\alpha$-decay rate of Po-210 nuclei follows the exponential decay law,
$n_{\rm Po}(t) \propto \exp(-t/\tau_{\rm Po})$. The deviations from this behaviour can be used
to determine the $\beta$-decay rate of Bi-210 nuclei. 
It was shown in Ref.\citenum{noiCNO} that a Borexino-like detector could
start discerning CNO neutrino signal in $\Delta t \sim 1 {\rm yr}$, if
the initial Po-210 event rate is $\sim 2000$cpd/100ton or lower. 
Future Kton-scale detectors, like e.g. SNO+, in the same time interval, could begin to discriminate between high and low metallicity solar models. 
The required assumptions are that the $\alpha$-particle detection
efficiency is stable and  external sources of Po-210 are negligible
during the data acquisition period. This requires stabilizing the
detector, avoiding in particular convective motions in the liquid
scintillator that may bring additional Po-210 in the fiducial volume, over time
scales comparable with the Polonium lifetime. The efforts of Borexino 
in this direction are described in D. Guffanti contribution to this
Workshop.

Finally, we discuss a component of the solar neutrino spectrum which
usually not included in solar neutrino analysis. As it was pointed out in Refs.\citenum{ecCNORob,ecCNOBah} and then considered
in Ref.\citenum{villanteecCNO}, along with neutrinos originating from the $\beta^+$ 
decay of $^{13}{\rm N}$, $^{15}{\rm O}$ and $^{17}{\rm F}$, 
neutrinos are also produced in the CNO cycle by the electron capture reactions:
\begin{eqnarray}
\nonumber
^{13}{\rm N}+e^-&\rightarrow&^{13}{\rm C}+\nu_e\\
^{15}{\rm O}+e^-&\rightarrow&^{15}{\rm N}+\nu_e\\
\nonumber
^{17}{\rm F}+e^-&\rightarrow&^{17}{\rm O}+\nu_e
\end{eqnarray}
The resulting fluxes, to which we refer as {\em ecCNO neutrino} fluxes, are
extremely small, at the level of 0.1\% with respect to the
“conventional” CNO neutrino fluxes. However, ecCNO neutrinos are
monochromatic and have larger energies equal to $E_\nu \sim 2.5\,{\rm MeV}$.
In Ref.\citenum{villanteecCNO}, it was suggested suggest that these
characteristics could make their detection possible in gigantic
ultra-pure liquid scintillator detectors.
The expected event rate is extremely low, at the level of few
counts/1kton/year in the observation window between $E_{\rm vis}\simeq [1.5\,
{\rm MeV}, 2.5\,{\rm MeV}]$ above the conventional CNO neutrinos endpoint, see
Ref.\citenum{villanteecCNO} for details. We thus understand that
detectors with fiducial masses equal to $\sim 10 \,{\rm kton}$ or more are
necessary, for statistical reasons, to extract the ecCNO neutrino
signal. 
Moreover, detectors should be placed underground at a depth comparable to
Pyhasalmi and SNOLAB, in order to prevent a too large cosmogenic background.
In conclusion, the determination of this sub-dominant component of the
solar neutrino flux is extremely difficult but could be rewarding in
terms of physical implications.
 Indeed, besides testing the efficiency of the CNO cycle and
 probing the metallic content of the solar core, it could also
 provide a determination of the electron neutrino survival probability
 at the energy $E_\nu\simeq 2.5\,{\rm MeV}$ which is otherwise inaccessible, with important implications for the final confirmation of the LMA-MSW flavour oscillation paradigm.

 \section{Summary}
\label{sec:summary}

In this review, we have described the B16-SSMs that includes recent updates on some important nuclear reaction rates,
a more consistent treatment of the equation of state and a novel and
flexible treatment of opacity uncertainties. 
We have quantified the level of agreement of the B16-SSMs calculated 
with two different canonical sets of solar abundances,
namely the (old, high metallicity) GS98 and the (new, low metallicity)
AGSS09met composition, with the helioseismic determination of the sound
speed profile, the depth of the convective envelope, the surface
helium abundance and with solar neutrino data. 

We have confirmed that solar models implementing the AGSS09met
abundances, based on three-dimensional hydrodynamical simulations of
the solar atmosphere (rather than the simplified one-dimensional
static models used in the past), do not reproduce helioseismic constraints. 
We have argued that this solar abundance problem can be equally solved by a
change of the composition and/or of the opacity of the solar plasma,
since effects produced by variations of metal abundances are
equivalent to those produced by suitable modifications of the solar
opacity profile, as it is quantitatively expressed by Eq.\ref{kappasource}.

Finally, we have emphasised the importance of neutrinos produced in the CNO
cycle, either the conventional CNO neutrinos produced by $\beta$-decay
of $^{13}{\rm N}$, $^{15}{\rm O}$ , $^{17}{\rm F}$ or the less abundant ecCNO neutrinos produced by electron capture reactions
on the same nuclei,  for removing the composition-opacity degeneracy 
and we have discussed the perspective for their future detection.

\bibliographystyle{ws-procs9x6}
\bibliography{villante}

\begin{thebibliography}{10}

\bibitem{Homestake1}
R.~Davis, Jr., D.~S. Harmer and K.~C. Hoffman, {Search for neutrinos from the
  sun}, {\em Phys. Rev. Lett.} {\bf 20}, 1205  (1968).

\bibitem{Homestake2}
B.~T. Cleveland, T.~Daily, R.~Davis, Jr., J.~R. Distel, K.~Lande, C.~K. Lee,
  P.~S. Wildenhain and J.~Ullman, {Measurement of the solar electron neutrino
  flux with the Homestake chlorine detector}, {\em Astrophys. J.} {\bf 496},
  505  (1998).

\bibitem{Gallex}
W.~Hampel {\em et~al.}, {GALLEX solar neutrino observations: Results for GALLEX
  IV}, {\em Phys. Lett.} {\bf B447}, 127  (1999).

\bibitem{SAGE}
J.~N. Abdurashitov {\em et~al.}, {Measurement of the solar neutrino capture
  rate by SAGE and implications for neutrino oscillations in vacuum}, {\em
  Phys. Rev. Lett.} {\bf 83}, 4686  (1999).

\bibitem{GNO}
M.~Altmann {\em et~al.}, {Complete results for five years of GNO solar neutrino
  observations}, {\em Phys. Lett.} {\bf B616}, 174  (2005).

\bibitem{Kamiokande}
K.~S. Hirata {\em et~al.}, {Observation of B-8 Solar Neutrinos in the
  Kamiokande-II Detector}, {\em Phys. Rev. Lett.} {\bf 63}, p.~16  (1989).

\bibitem{SK}
J.~P. Cravens {\em et~al.}, {Solar neutrino measurements in
  Super-Kamiokande-II}, {\em Phys. Rev.} {\bf D78}, p. 032002  (2008).

\bibitem{SNO}
Q.~R. Ahmad {\em et~al.}, {Measurement of day and night neutrino energy spectra
  at SNO and constraints on neutrino mixing parameters}, {\em Phys. Rev. Lett.}
  {\bf 89}, p. 011302  (2002).

\bibitem{SKLatest}
K.~Abe {\em et~al.}, {Solar Neutrino Measurements in Super-Kamiokande-IV}, {\em
  Phys. Rev.} {\bf D94}, p. 052010  (2016).

\bibitem{SNOLatest}
A.~Bellerive, J.~R. Klein, A.~B. McDonald, A.~J. Noble and A.~W.~P. Poon, {The
  Sudbury Neutrino Observatory}, {\em Nucl. Phys.} {\bf B908}, 30  (2016).

\bibitem{BorexPP}
G.~Bellini {\em et~al.}, {Neutrinos from the primary proton–proton fusion
  process in the Sun}, {\em Nature} {\bf 512}, 383  (2014).

\bibitem{BorexLatest}
M.~Agostini {\em et~al.}, {First Simultaneous Precision Spectroscopy of $pp$,
  $^7$Be, and $pep$ Solar Neutrinos with Borexino Phase-II}  (2017).

\bibitem{Borex8BLatest}
M.~Agostini {\em et~al.}, {Improved measurement of $^8$B solar neutrinos with
  1.5 kt y of Borexino exposure}  (2017).

\bibitem{noiCNO}
F.~L. Villante, A.~Ianni, F.~Lombardi, G.~Pagliaroli and F.~Vissani, {A Step
  toward CNO solar neutrinos detection in liquid scintillators}, {\em Phys.
  Lett.} {\bf B701}, 336  (2011).

\bibitem{HelioReview1}
S.~Degl'Innocenti, W.~A. Dziembowski, G.~Fiorentini and B.~Ricci,
  {Helioseismology and standard solar models}, {\em Astropart. Phys.} {\bf 7},
  77  (1997).

\bibitem{HelioReview2}
D.~O. Gough {\em et~al.}, {The seismic structure of the Sun}, {\em Science}
  {\bf 272}, 1296  (1996).

\bibitem{agss09}
M.~Asplund, N.~Grevesse, A.~J. Sauval and P.~Scott, {The chemical composition
  of the Sun}, {\em Ann. Rev. Astron. Astrophys.} {\bf 47}, 481  (2009).

\bibitem{caffau}
E.~Caffau, H.-G. Ludwig, M.~Steffen, B.~Freytag and P.~Bonifacio, {Solar
  Chemical Abundances Determined with a CO5BOLD 3D Model Atmosphere}, {\em
  Solar Phys.} {\bf 268}, p. 255  (2011).

\bibitem{ags05}
M.~{Asplund}, N.~{Grevesse} and A.~J. {Sauval}, {The Solar Chemical
  Composition}, in {\em Cosmic Abundances as Records of Stellar Evolution and
  Nucleosynthesis\/},  eds. T.~G. {Barnes}, III and F.~N. {Bash}, Astronomical
  Society of the Pacific Conference Series, Vol.~336September 2005.

\bibitem{gs98}
N.~Grevesse and A.~J. Sauval, {Standard Solar Composition}, {\em Space Sci.
  Rev.} {\bf 85}, 161  (1998).

\bibitem{gn93}
N.~{Grevesse} and A.~{Noels}, {Cosmic abundances of the elements.}, in {\em
  Origin and Evolution of the Elements\/},  eds. N.~{Prantzos},
  E.~{Vangioni-Flam} and M.~{Casse}January 1993.

\bibitem{serenelli14}
M.~{Bergemann} and A.~{Serenelli}, {\em {Solar Abundance Problem}}, in {\em
  Determination of Atmospheric Parameters of B-, A-, F- and G-Type
  Stars.~Series: GeoPlanet: Earth and Planetary Sciences, ISBN:
  978-3-319-06955-5. Springer International Publishing (Cham), Edited by Ewa
  Niemczura, Barry Smalley and Wojtek Pych, pp.~245-258\/},  eds.
  E.~{Niemczura}, B.~{Smalley} and W.~{Pych} 2014, pp. 245--258.

\bibitem{Basu:2007fp}
S.~Basu and H.~M. Antia, {Helioseismology and Solar Abundances}, {\em Phys.
  Rept.} {\bf 457}, 217  (2008).

\bibitem{serenelli11}
A.~M. Serenelli, W.~C. Haxton and C.~Pena-Garay, {Solar models with accretion.
  I. Application to the solar abundance problem}, {\em Astrophys. J.} {\bf
  743}, p.~24  (2011).

\bibitem{freeEOS}
S.~Cassisi, M.~Salaris and A.~W. Irwin, {The initial helium content of galactic
  globular cluster stars from the r-parameter: comparison with the cmb
  constraint}, {\em Astrophys. J.} {\bf 588}, p. 862  (2003).

\bibitem{noi}
N.~Vinyoles, A.~M. Serenelli, F.~L. Villante, S.~Basu, J.~Bergström, M.~C.
  Gonzalez-Garcia, M.~Maltoni, C.~Peña-Garay and N.~Song, {A new Generation of
  Standard Solar Models}, {\em Astrophys. J.} {\bf 835}, p. 202  (2017).

\bibitem{deboer}
R.~J. deBoer, J.~Görres, K.~Smith, E.~Uberseder, M.~Wiescher, A.~Kontos,
  G.~Imbriani, A.~Di~Leva and F.~Strieder, {Monte Carlo uncertainty of the
  He3($\alpha$,gamma)Be7 reaction rate}, {\em Phys. Rev.} {\bf C90}, p. 035804
  (2014).

\bibitem{iliadis}
C.~Iliadis, K.~Anderson, A.~Coc, F.~Timmes and S.~Starrfield, {Bayesian
  Estimation of Thermonuclear Reaction Rates}, {\em Astrophys. J.} {\bf 831},
  p. 107  (2016).

\bibitem{adelberger11}
E.~G. Adelberger {\em et~al.}, {Solar fusion cross sections II: the pp chain
  and CNO cycles}, {\em Rev. Mod. Phys.} {\bf 83}, p. 195  (2011).

\bibitem{s11a}
L.~E. Marcucci, R.~Schiavilla and M.~Viviani, {Proton-Proton Weak Capture in
  Chiral Effective Field Theory}, {\em Phys. Rev. Lett.} {\bf 110}, p. 192503
  (2013).

\bibitem{s11b}
E.~Tognelli, S.~Degl'Innocenti, L.~E. Marcucci and P.~G. Prada~Moroni,
  {Astrophysical implications of the proton–proton cross section updates},
  {\em Phys. Lett.} {\bf B742}, 189  (2015).

\bibitem{s11c}
B.~Acharya, B.~D. Carlsson, A.~Ekström, C.~Forssén and L.~Platter,
  {Uncertainty quantification for proton–proton fusion in chiral effective
  field theory}, {\em Phys. Lett.} {\bf B760}, 584  (2016).

\bibitem{s17}
X.~Zhang, K.~M. Nollett and D.~R. Phillips, {Halo effective field theory
  constrains the solar $^7$Be + p $\to$ $^8$B + $\gamma$ rate}, {\em Phys.
  Lett.} {\bf B751}, 535  (2015).

\bibitem{s114}
M.~Marta {\em et~al.}, {The $^{14}$N(p,$\gamma$)$^{15}$O reaction studied with
  a composite germanium detector}, {\em Phys. Rev.} {\bf C83}, p. 045804
  (2011).

\bibitem{OP}
N.~R. Badnell, M.~A. Bautista, K.~Butler, F.~Delahaye, C.~Mendoza, P.~Palmeri,
  C.~J. Zeippen and M.~J. Seaton, {Up-dated opacities from the Opacity
  Project}, {\em Mon. Not. Roy. Astron. Soc.} {\bf 360}, 458  (2005).

\bibitem{OPAL}
C.~A. Iglesias and F.~J. Rogers, {Updated Opal Opacities}, {\em Astrophys. J.}
  {\bf 464}, p. 943  (1996).

\bibitem{villanteOPA}
F.~L. Villante, {Constraints on the opacity profile of the sun from
  helioseismic observables and solar neutrino flux measurements}, {\em
  Astrophys. J.} {\bf 724}, 98  (2010).

\bibitem{OPAexp}
J.~E. {Bailey}, T.~{Nagayama}, G.~P. {Loisel}, G.~A. {Rochau}, C.~{Blancard},
  J.~{Colgan}, P.~{Cosse}, G.~{Faussurier}, C.~J. {Fontes}, F.~{Gilleron},
  I.~{Golovkin}, S.~B. {Hansen}, C.~A. {Iglesias}, D.~P. {Kilcrease}, J.~J.
  {Macfarlane}, R.~C. {Mancini}, S.~N. {Nahar}, C.~{Orban}, J.-C. {Pain}, A.~K.
  {Pradhan}, M.~{Sherrill} and B.~G. {Wilson}, {A higher-than-predicted
  measurement of iron opacity at solar interior temperatures}, {\em Nature}
  {\bf 517}, 56 (January 2015).

\bibitem{OPAtheo}
M.~Krief, A.~Feigel and D.~Gazit, {Solar opacity calculations using the
  super-transition-array method}, {\em Astrophys. J.} {\bf 821}, p.~45  (2016).

\bibitem{agss15a}
P.~Scott, N.~Grevesse, M.~Asplund, A.~J. Sauval, K.~Lind, Y.~Takeda, R.~Collet,
  R.~Trampedach and W.~Hayek, {The elemental composition of the Sun I. The
  intermediate mass elements Na to Ca}, {\em Astron. Astrophys.} {\bf 573}, p.
  A25  (2015).

\bibitem{agss15b}
P.~Scott, M.~Asplund, N.~Grevesse, M.~Bergemann and A.~J. Sauval, {The
  elemental composition of the Sun II. The iron group elements Sc to Ni}, {\em
  Astron. Astrophys.} {\bf 573}, p. A26  (2015).

\bibitem{agss15c}
N.~Grevesse, P.~Scott, M.~Asplund and A.~J. Sauval, {The elemental composition
  of the Sun III. The heavy elements Cu to Th}, {\em Astron. Astrophys.} {\bf
  573}, p. A27  (2015).

\bibitem{bergstrom}
J.~Bergstrom, M.~C. Gonzalez-Garcia, M.~Maltoni, C.~Pena-Garay, A.~M. Serenelli
  and N.~Song, {Updated determination of the solar neutrino fluxes from solar
  neutrino data}, {\em JHEP} {\bf 03}, p. 132  (2016).

\bibitem{YsErr}
S.~Basu and H.~M. Antia, {Constraining solar abundances using helioseismology},
  {\em Astrophys. J.} {\bf 606}, p. L85  (2004).

\bibitem{RczErr}
S.~{Basu} and H.~M. {Antia}, {Seismic measurement of the depth of the solar
  convection zone}, {\em Mon. Not. Roy. Astron. Soc.} {\bf 287}, 189 (May
  1997).

\bibitem{villanteCC}
F.~L. Villante, A.~M. Serenelli, F.~Delahaye and M.~H. Pinsonneault, {The
  chemical composition of the Sun from helioseismic and solar neutrino data},
  {\em Astrophys. J.} {\bf 787}, p.~13  (2014).

\bibitem{BISON}
S.~Basu, W.~J. Chaplim, Y.~Elsworth, R.~New and A.~M. Serenelli, {Fresh
  insights on the structure of the solar core}, {\em Astrophys. J.} {\bf 699},
  1403  (2009).

\bibitem{villanteLSM}
F.~L. Villante and B.~Ricci, {Linear Solar Models}, {\em Astrophys. J.} {\bf
  714}, 944  (2010).

\bibitem{villanteWIMPs}
A.~Bottino, G.~Fiorentini, N.~Fornengo, B.~Ricci, S.~Scopel and F.~L. Villante,
  {Does solar physics provide constraints to weakly interacting massive
  particles?}, {\em Phys. Rev.} {\bf D66}, p. 053005  (2002).

\bibitem{JCD}
S.~C. Tripathy and J.~Christensen-Dalsgaard, {Opacity effects on the solar
  interior. I. solar structure}, {\em Astron. Astrophys.} {\bf 337}, p. 579
  (1998).

\bibitem{SerenelliHaxton}
W.~C. Haxton and A.~M. Serenelli, {CN-Cycle Solar Neutrinos and Sun's
  Primordial Core Metalicity}, {\em Astrophys. J.} {\bf 687}, 678  (2008).

\bibitem{BorexBiCNO}
C.~Arpesella {\em et~al.}, {First real time detection of Be-7 solar neutrinos
  by Borexino}, {\em Phys. Lett.} {\bf B658}, 101  (2008).

\bibitem{PRLBorex}
C.~Arpesella {\em et~al.}, {Direct Measurement of the Be-7 Solar Neutrino Flux
  with 192 Days of Borexino Data}, {\em Phys. Rev. Lett.} {\bf 101}, p. 091302
  (2008).

\bibitem{ecCNORob}
L.~C. Stonehill, J.~A. Formaggio and R.~G.~H. Robertson, {Solar neutrinos from
  CNO electron capture}, {\em Phys. Rev.} {\bf C69}, p. 015801  (2004).

\bibitem{ecCNOBah}
J.~N. Bahcall, {Line versus continuum solar neutrinos}, {\em Phys. Rev.} {\bf
  D41}, p. 2964  (1990).

\bibitem{villanteecCNO}
F.~L. Villante, {ecCNO Solar Neutrinos: A Challenge for Gigantic Ultra-Pure
  Liquid Scintillator Detectors}, {\em Phys. Lett.} {\bf B742}, 279  (2015).

\end{thebibliography}

\end{document}